# Optimizing scheme of probabilistic remote preparation state

Xin-wei Zha, Jia-fan Xia, Jian-xia Qi

*School of Science, Xi'an University of Posts and Telecommunications, Xi'an, 710121, P R China*

*E-mail: zhxw@xupt.edu.cn*

**Abstract**

We propose a new generalized remote state preparation protocol for using non-maximally entangled state as a shared resource. Different from the previous schemes, the parameters of measurement basis depend on not only the state of preparation but also the state of quantum channel. The advantage of the present protocols is that, by constructing a set of appropriate mutually orthogonal measurement basis, the success probability is improved. It is worthwhile noticing that the probability of success is determined by Alice's measurement. If Alice's measurement is appropriate, the remote preparation can be successfully realized with the maximal probability.

## 1. Introduction

Quantum entanglement is a valuable resource for the implementation of quantum computation and quantum communication protocols, like quantum teleportation [1], remote state preparation [2,3] and so on. If a sender (Alice) wants to transmit an unknown quantum state to a receiver (Bob), she may use teleportation. In remote state preparation (RSP), Alice is assumed to know entirely the transmitted state to be prepared by Bob, however in the process of teleportation neither Alice nor Bob know something about the transmitted state. In the last decade, RSP has already attracted much attention [4–23]. Especially, some probabilistic remote state preparation schemes using partial entangled state as the quantum channel have been proposed [24–30].

However, in all remote state preparation protocol, whether using maximally entangled state or non-maximally entangled state as the quantum channel, the Alice's measurement basis is same. But we know that RSP is different to quantum teleportation. In RSP, the sender Alice can perform some special chosen measurements on her particle according to the original state that she has known. Therefore，different from the previous scheme, we propose a new scheme of remote preparation states using a partial entangled states as the quantum channel. In which the parameter of measurement basis depends on not only the state of preparation but also the state of quantum channel. We have shown in general that the proposed protocol is superior to the existing ones. And the significant advantage is that the success probability for remote preparing states is higher than others.

## 2．Remote preparation of an arbitrary single-qubit state

We firstly present the remote preparation of a single-qubit state with real coefficients and complex coefficients, respectively. In the protocol a non-maximally entangled state is sued as quantum channel.

### 2.1 RSP of a single-qubit state with real coefficients

Suppose that Alice wishes to help Bob to prepare remotely the state

$$|\varphi\rangle = \alpha|0\rangle + \beta|1\rangle \tag{1}$$

where $\alpha, \beta$ are real numbers and satisfy the normalized condition $\alpha^2 + \beta^2 = 1$. And we assume that $\alpha \leq \beta$.

Assuming that Alice knows this state precisely, which means she knows $\alpha, \beta$ completely, but Bob does not know them at all. We set that the quantum channel shared by Alice and Bob is a non-maximally entangled states, i.e.,

$$|\varphi\rangle_{AB} = (a|00\rangle + b|11\rangle)_{AB}, \tag{2}$$

where the coefficients a and b are all nonzero real numbers and satisfy $a^2 + b^2 = 1$. And we make $a \leq b$, and particle A belongs to Alice, while particle B belongs to Bob.

In order to help Bob remotely prepare the original state (1), what Alice need to do is to perform a single-qubit projective measurement on her qubit $A$, the RSP protocol can be probabilistically achieved. We find that Alice's measurement basis can be described by a set of mutually orthogonal basis vectors (MOBVs)

$$|\varphi^1\rangle_A = c_1(\alpha b|0\rangle + \beta a|1\rangle)_A,$$
$$|\varphi^2\rangle_A = c_1(\beta a|0\rangle - \alpha b|1\rangle)_A, \tag{3}$$

where $c_1 = \dfrac{1}{\sqrt{b^2\alpha^2 + a^2\beta^2}}$.

When Bob is informed the actual measurement outcome by Alice via a classical channel, he can get the original state described in Eq.(1) with certain probability. Without loss of generality, if Alice's measurement result is $|\varphi^1\rangle_A$, the state of qubit B has been collapsed into

$$|\varphi^1\rangle_B = c_1 ab(\alpha|0\rangle + \beta|1\rangle)_B \tag{4}$$

Which is exactly the original state $|\varphi\rangle$, and that the successful probability is $(c_1 ab)^2$.

If Alice's measurement result is $|\varphi^2\rangle_A$, the state of qubit B has been collapsed into

$$|\varphi^2\rangle_B = c_1(a^2\beta|0\rangle - b^2\alpha|1\rangle)_B \tag{5}$$

and that the probability is $c_1^2(a^4\beta^2 + b^4\alpha^2)$

In order to reconstruct the original state at his side, Bob introduces an auxiliary two-level particle b with the initial state $|0\rangle_b$ and performs a collective unitary transformation $U_{Bb}$ on his particle B and b under the basis $\{|00\rangle_{Bb}, |10\rangle_{Bb}, |01\rangle_{Bb}, |11\rangle_{Bb}\}$. And the unitary $U_{Bb}$ can be

expressed as

$$U_{Bb} = \begin{pmatrix} \mathcal{F}_1 & \mathcal{G}_1 \\ \mathcal{G}_1 & -\mathcal{F}_1 \end{pmatrix}_{4\times 4}, \tag{6}$$

where

$$\mathcal{F}_1 = diag\left(1, \frac{a^2}{b^2}\right), \tag{7}$$

and

$$\mathcal{G}_1 = diag\left(0, \sqrt{1-\left(\frac{a^2}{b^2}\right)^2}\right). \tag{8}$$

The unitary transformation $U_{Bb}$ will transform $|\varphi^2\rangle_B |0\rangle_b$ into

$$c_1[a^2(\beta|0\rangle - \alpha|1\rangle)_B |0\rangle_b + \sqrt{b^4 - a^4}\,\alpha|1\rangle_B |1\rangle_b] \tag{9}$$

Then Bob measures the state of auxiliary particle $b$. If his measurement result is $|1\rangle_b$, the remote preparation fails. If the result is $|0\rangle_b$, the state of particle B collapses to $c_1 a^2(\beta|0\rangle - \alpha|1\rangle)_B$.

Hence, the remote preparation of the original state is successfully realized and the probability that Bob obtains the state $|0\rangle_b$ is $a^4/(a^4\beta^2 + b^4\alpha^2)$. And the probability of successful RSP for Alice's von Neumann measurement result $|\varphi^2\rangle_B$ can be gotten. It is $(c_1 a^2)^2$.

By analyzing, we can get the probability of successful RSP. It equals to the sum of the above measurement. That is, it can be written as

$$(c_1 ab)^2 + (c_1 a^2)^2 = (c_1 a)^2 = \frac{a^2}{b^2\alpha^2 + a^2\beta^2} \tag{10}$$

Because of $\alpha \leq \beta$, $|a| < |b|$, we can have the following result $b^2\alpha^2 + a^2\beta^2 \leq \frac{1}{2}$.

Therefore, the probability of successful RSP is greater than or equal to $2a^2$. That is

$$\frac{a^2}{b^2\alpha^2 + a^2\beta^2} \geq 2a^2 \tag{11}$$

Clearly, if $a=b$, the probability of successful RSP $\frac{a^2}{b^2\alpha^2+a^2\beta^2}=\frac{1}{\alpha^2+\beta^2}=1$, i.e. the probability of successful RSP is 1 with maximally entangled state quantum channel.

From the above analyses, we can find that the total probability of successful RSP depend on not only the parameter of quantum channel but also the parameter of the coefficients in Eq. (1).

**2.2 RSP of a single-qubit state with complex coefficients**

Now let us further consider the general case of the remote preparation of a single-qubit state. Suppose that the state Alice wishes to help Bob remotely prepare is

$$|\varphi\rangle = \alpha|0\rangle + \beta e^{i\varphi}|1\rangle \tag{12}$$

where $\alpha$, $\beta, \varphi$ are real numbers and satisfy the normalized condition $\alpha^2+\beta^2=1$. And we assume that $\alpha \leq \beta$. We set the quantum channel shared by Alice and Bob is a non-maximally entangled states, i.e.,

$$|\varphi\rangle_{AB} = (a|00\rangle + b|11\rangle)_{AB}, \tag{13}$$

where the coefficients a and b are all nonzero complex numbers and satisfy $|a|^2+|b|^2=1$, And assuming that $|a|<|b|$. In the states, particle A belongs to Alice, while particle B belongs to Bob. About this case, the Alice's measurement basis can be described

$$|\varphi^1\rangle_A = c_2\left(\alpha b^*|0\rangle + \beta e^{-i\varphi}a^*|1\rangle\right)_A,$$
$$|\varphi^2\rangle_A = c_2\left(\beta e^{i\varphi}a|0\rangle - \alpha b^*|1\rangle\right)_A, \tag{14}$$

where $c_2 = \frac{1}{\sqrt{|b|^2\alpha^2+|a|^2\beta^2}}$.

If Alice's measurement result is $|\varphi^1\rangle_A$, the state of qubit B has been collapsed into

$$|\varphi^1\rangle_B = c_2 ab\left(\alpha|0\rangle + \beta e^{i\varphi}|1\rangle\right)_B \tag{15}$$

Which is exactly the original state $|\varphi\rangle$, and that the successful probability is $c_2^2|ab|^2$. However, when the measurement outcome is $|\varphi^2\rangle_A$, the remote state preparation can not be successful, Therefore, we can get the probability of successful RSP is

$$c_2^2|ab|^2 = \frac{|ab|^2}{|b|^2\alpha^2+|a|^2\beta^2} \tag{16}$$

Because of $\alpha \leq \beta$, $|a| < |b|$, we can obtain the following result $|b|^2 \alpha^2 + |a|^2 \beta^2 \leq \frac{1}{2}$. That is

$$\frac{|ab|^2}{|b|^2 \alpha^2 + |a|^2 \beta^2} \geq 2|ab|^2 \tag{17}$$

obviously, if $|a|=|b|$, the probability of successful RSP $\frac{|ab|^2}{|b|^2 \alpha^2 + |a|^2 \beta^2} = \frac{1}{2}$, i.e. the probability of successful RSP is 1/2 with maximally entangled state quantum channel.

## 3. Remote preparation of an arbitrary two-qubit state

Now let us further consider the remote preparation of an arbitrary two-qubit state with real coefficients and complex coefficients, respectively.

### 3.1 RSP of a two-qubit state with real coefficients

Suppose that the sender Alice wants to help a remote receiver Bob prepare the following two-particle state

$$|\varphi\rangle_a = \alpha|00\rangle + \beta|01\rangle + \gamma|10\rangle + \delta|11\rangle \tag{18}$$

where α, β, γ and δ are real number and $\alpha^2 + \beta^2 + \gamma^2 + \delta^2 = 1$. We assume that $\alpha \leq \beta \leq \gamma \leq \delta$.

The quantum channels shared by Alice and Bob are non-maximally entangled states

$$|\varphi\rangle_{A_1 A_2 B_1 B_2} = (a|0000\rangle + b|0101\rangle + c|1010\rangle + d|1111\rangle)_{A_1 A_2 B_2 B_1} \tag{19}$$

Where $a^2 + b^2 + c^2 + d^2 = 1$, Supposing $a \leq b \leq c \leq d$, $ad \leq bc$. Particles $A_1, A_2$ belong to Alice, while particles $B_1, B_2$ belong to Bob. The measurement basis chosen by Alice is a set of mutually orthogonal basis vectors $\{|\varphi^1\rangle_{A_1 A_2}, |\varphi^2\rangle_{A_1 A_2}, |\varphi^3\rangle_{A_1 A_2}, |\varphi^4\rangle_{A_1 A_2}\}$. We find that this basis can be related to the computation basis $\{|00\rangle_{A_1 A_2}, |01\rangle_{A_1 A_2}, |10\rangle_{A_1 A_2}, |11\rangle_{A_1 A_2}\}$ by the following relations:

$$|\varphi^1\rangle_{A_1 A_2} = k\left(bcd\alpha|00\rangle + adc\beta|01\rangle + abd\gamma|10\rangle + abc\delta|11\rangle\right)_{A_1 A_2},$$

$$|\varphi^2\rangle_{A_1 A_2} = k\left(adc\beta|00\rangle - bcd\alpha|01\rangle - abc\delta|10\rangle + abd\gamma|11\rangle\right)_{A_1 A_2},$$

$$|\varphi^3\rangle_{A_1 A_2} = k\left(abd\gamma|00\rangle + abc\delta|01\rangle - bcd\alpha|10\rangle - adc\beta|11\rangle\right)_{A_1 A_2},$$

$$|\varphi^4\rangle_{A_1 A_2} = k\left(-abc\delta|00\rangle + abd\gamma|01\rangle - adc\beta|10\rangle + bcd\alpha|11\rangle\right)_{A_1 A_2}, \tag{20}$$

where $k = \frac{1}{\sqrt{b^2 c^2 d^2 \alpha^2 + a^2 c^2 d^2 \beta^2 + a^2 b^2 d^2 \gamma^2 + a^2 b^2 c^2 \delta^2}}$.

Therefore, we can have the following expression,

$$|\varphi\rangle_{A_1A_2B_1B_2} = [|\varphi^1\rangle_{A_1A_2}|\varphi^1\rangle_{B_1B_2} + |\varphi^2\rangle_{A_1A_2}|\varphi^2\rangle_{B_1B_2} + |\varphi^3\rangle_{A_1A_2}|\varphi^3\rangle_{B_1B_2} + |\varphi^4\rangle_{A_1A_2}|\varphi^4\rangle_{B_1B_2}]. \tag{21}$$

From Eqs.(20-21), it is easy to obtain

$$|\varphi^1\rangle_{B_1B_2} = kabcd(\alpha|00\rangle + \beta|01\rangle + \gamma|10\rangle + \delta|11\rangle)_{B_1B_2},$$

$$|\varphi^2\rangle_{B_1B_2} = k(a^2cd\beta|00\rangle - b^2cd\alpha|01\rangle - abc^2\delta|10\rangle + abd^2\gamma|11\rangle)_{B_1B_2},$$

$$|\varphi^3\rangle_{B_1B_2} = k(a^2bd\gamma|00\rangle + ab^2c\delta|01\rangle - bc^2d\alpha|10\rangle - acd^2\beta|11\rangle)_{B_1B_2},$$

$$|\varphi^4\rangle_{B_1B_2} = k(-a^2bc\delta|00\rangle + ab^2d\gamma|01\rangle - ac^2d\beta|10\rangle + bcd^2\alpha|11\rangle)_{B_1B_2} \tag{22}$$

As we know, if Alice's von Neumann measurement result is $|\varphi^i\rangle_{A_1A_2}$ $(i=1,2,3,4)$, the state of particles $B_1, B_2$ belonging to Bob, will collapse into $|\varphi^i\rangle_{B_1B_2}$ $(i=1,2,3,4)$.

For example, if Alice's von Neumann measurement result is $|\varphi^1\rangle_{B_1B_2}$, the state of particles $B_1, B_2$ as shown by Eq. (22) will collapse into

$$|\varphi^1\rangle_{B_1B_2} = kabcd(\alpha|00\rangle + \beta|01\rangle + \gamma|10\rangle + \delta|11\rangle)_{B_1B_2} \tag{23}$$

Which is exactly the original state $|\varphi\rangle$, and that the successful probability is $(kabcd)^2$.

If Alice's von Neumann measurement result is $|\varphi^2\rangle_{B_1B_2}$, the state of particles $B_1, B_2$ will collapse into

$$|\varphi^2\rangle_{B_1B_2} = k(a^2cd\beta|00\rangle - b^2cd\alpha|01\rangle - abc^2\delta|10\rangle + abd^2\gamma|11\rangle)_{B_1B_2}, \tag{24}$$

In order to recover the original state, Bob introduces an auxiliary two-level particle $b$ with the initial state $|0\rangle_b$ and performs a collective unitary transformation $U^1_{B_1B_2b}$ on his particles $B_1, B_2$ and b under the basis

$\{|000\rangle_{B_1B_2b}, |010\rangle_{B_1B_2b}, |100\rangle_{B_1B_2b}, |110\rangle_{B_1B_2b}, |001\rangle_{B_1B_2b}, |011\rangle_{B_1B_2b}, |101\rangle_{B_1B_2b}, |111\rangle_{B_1B_2b}\}$. And the unitary $U^1_{B_1B_2b}$ can be expressed as

$$U^1_{B_1B_2b} = \begin{pmatrix} \mathcal{F}_1 & \mathcal{G}_1 \\ \mathcal{G}_1 & -\mathcal{F}_1 \end{pmatrix}_{8\times 8}, \tag{25}$$

where

$$\mathcal{F}_1 = diag\left(1, \frac{a^2}{b^2}, \frac{ad}{bc}, \frac{bd}{ac}\right), \tag{26}$$

and

$$\mathcal{G}_1 = diag\left(0, \sqrt{1-\left(\frac{a^2}{b^2}\right)^2}, \sqrt{1-\left(\frac{ad}{bc}\right)^2}, \sqrt{1-\left(\frac{bd}{ac}\right)^2}\right). \tag{27}$$

The unitary transformation $U^1_{B_1B_2b}$ will transform $|\varphi^2\rangle_{B_1B_2}|0\rangle_b$ into

$$|\varphi^2\rangle_{B_1B_2b} = ka^2cd[(\beta|00\rangle - \alpha|01\rangle - \delta|10\rangle + \gamma|11\rangle)_{B_1B_2}|0\rangle_b$$
$$+(-cd\sqrt{b^4-a^4}\alpha|01\rangle_{B_1B_2} - ac\sqrt{b^2c^2-a^2d^2}\delta|10\rangle_{B_1B_2} + ad\sqrt{b^2d^2-a^2c^2}\gamma|11\rangle_{B_1B_2})|1\rangle_b] \tag{28}$$

Then Bob measures the state of auxiliary particle $b$. If his measurement result is $|1\rangle_b$, the remote preparation fails. If the result is $|0\rangle_b$, the state of particles $B_1, B_2$ collapses to

$$ka^2cd\left(\beta|00\rangle - \alpha|01\rangle - \delta|10\rangle + \gamma|11\rangle\right) \tag{29}$$

Hence, the remote preparation of the original state is successfully realized and that the successful probability is $(ka^2cd)^2$.

Similarly, If Alice's von Neumann measurement result is $|\varphi^3\rangle_{B_1B_2}$, the unitary $U^2_{B_1B_2b}$ can be expressed as

$$U^2_{B_1B_2b} = \begin{pmatrix} \mathcal{F}_2 & \mathcal{G}_2 \\ \mathcal{G}_2 & -\mathcal{F}_2 \end{pmatrix}_{8\times 8}, \tag{30}$$

where

$$\mathcal{F}_2 = diag\left(1, \frac{ad}{bc}, \frac{a^2}{c^2}, \frac{ab}{cd}\right), \tag{31}$$

and

$$\mathcal{G}_2 = diag\left(0, \sqrt{1-\left(\frac{ad}{bc}\right)^2}, \sqrt{1-\left(\frac{a^2}{c^2}\right)^2}, \sqrt{1-\left(\frac{ab}{cd}\right)^2}\right).$$

$$\tag{32}$$

Then Bob measures the state of auxiliary particle b. If his measurement result is $|1\rangle_b$, the remote preparation fails. If the result is $|0\rangle_b$, the remote preparation of the original state is successfully realized. And the successful probability is $(ka^2bd)^2$.

Also, If Alice's von Neumann measurement result is $|\varphi^4\rangle_{B_1B_2}$, the unitary $U^3_{B_1B_2b}$ can be expressed as

$$U^3_{B_1B_2b} = \begin{pmatrix} \mathcal{F}_3 & \mathcal{G}_3 \\ \mathcal{G}_3 & -\mathcal{F}_3 \end{pmatrix}_{8\times 8}, \qquad (33)$$

where

$$\mathcal{F}_3 = diag\left(1, \frac{ac}{bd}, \frac{ab}{cd}, \frac{a^2}{d^2},\right), \qquad (34)$$

and

$$\mathcal{G}_3 = diag\left(0, \sqrt{1-\left(\frac{ac}{bd}\right)^2}, \sqrt{1-\left(\frac{ab}{cd}\right)^2}, \sqrt{1-\left(\frac{a^2}{d^2}\right)^2}\right).$$

(35)

Then Bob measures the state of auxiliary particle b. If his measurement result is $|1\rangle_b$, the remote preparation fails. If the result is $|0\rangle_b$, the remote preparation of the original state is successfully realized. And the successful probability is $(ka^2bc)^2$

Thus, we can find that the total successful probability is

$$\begin{aligned}&(kabcd)^2 + (ka^2cd)^2 + (kbda^2)^2 + (kbca^2)^2 \\ &= \frac{a^2[c^2d^2(b^2+a^2)+a^2b^2(d^2+c^2)]}{c^2d^2(b^2\alpha^2+a^2\beta^2)+a^2b^2(d^2\gamma^2+c^2\delta^2)}\end{aligned} \qquad (36)$$

Because of $\alpha \leq \beta \leq \gamma \leq \delta$, $a \leq b \leq c \leq d$, we can have the following result

$$\frac{a^2[c^2d^2(b^2+a^2)+a^2b^2(d^2+c^2)]}{c^2d^2(b^2\alpha^2+a^2\beta^2)+a^2b^2(d^2\gamma^2+c^2\delta^2)} \geq 4a^2 \qquad (37)$$

obviously, if $a=b=c=d$, the probability of successful RSP

$$\frac{a^2[c^2d^2(b^2+a^2)+a^2b^2(d^2+c^2)]}{c^2d^2(b^2\alpha^2+a^2\beta^2)+a^2b^2(d^2\gamma^2+c^2\delta^2)} = 1,$$

i.e. the probability of successful RSP is 1 with maximally entangled state quantum channel.

## 3.2 RSP of a two-qbit sate with complex coefficients

Suppose that the state Alice wishes to help Bob remotely prepare is

$$|\varphi\rangle = \alpha|00\rangle + \beta e^{i\varphi_1}|01\rangle + \gamma e^{i\varphi_2}|10\rangle + \delta e^{i\varphi_3}|11\rangle \qquad (38)$$

where $\alpha, \beta, \gamma, \delta, \varphi_i$ are real numbers and satisfy the normalized condition $\alpha^2 + \beta^2 + \gamma^2 + \delta^2 = 1$. And we assume that $\alpha \leq \beta \leq \gamma \leq \delta$. The quantum channels shared by Alice and Bob are non-maximally entangled states

$$|\varphi\rangle_{A_1A_2B_1B_2} = (a|0000\rangle + b|0101\rangle + c|1010\rangle + d|1111\rangle)_{A_1A_2B_2B_1} \qquad (39)$$

where the coefficients a, b, c and d are all nonzero complex numbers and satisfy $|a|^2 + |b|^2 + |c|^2 + |d|^2 = 1$, And we assume that $|a| \leq |b| \leq |c| \leq |d|$, and particles $A_1, A_2$ belong to Alice, while particles $B_1, B_2$ belong to Bob.

Alice must make a measurement on her two particles $A_1, A_2$. The measurement basis chosen by Alice is a set of mutually orthogonal basis vectors $\{|\varphi^1\rangle_{A_1A_2}, |\varphi^2\rangle_{A_1A_2}, |\varphi^3\rangle_{A_1A_2}, |\varphi^4\rangle_{A_1A_2}\}$. We find that this basis can be related to the computation basis $\{|00\rangle_{A_1A_2}, |01\rangle_{A_1A_2}, |10\rangle_{A_1A_2}, |11\rangle_{A_1A_2}\}$ by the following relations:

$$|\varphi^1\rangle_{A_1A_2} = k\left(b^*c^*d^*\alpha|00\rangle + a^*c^*d^*\beta e^{-i\varphi_1}|01\rangle + a^*b^*d^*\gamma e^{-i\varphi_2}|10\rangle + a^*b^*c^*\delta e^{-i\varphi_3}|11\rangle\right)_{A_1A_2},$$

$$|\varphi^2\rangle_{A_1A_2} = k\left(adc\beta|00\rangle - bcd\alpha e^{-i\varphi_1}|01\rangle - abc\delta e^{-i\varphi_2}|10\rangle + abd\gamma e^{-i\varphi_3}|11\rangle\right)_{A_1A_2},$$

$$|\varphi^3\rangle_{A_1A_2} = k[\sqrt{\frac{|abd\gamma|^2 + |abc\delta|^2}{|bcd\alpha|^2 + |acd\beta|^2}}\left(b^*c^*d^*\alpha|00\rangle + a^*c^*d^*\beta e^{-i\varphi_1}|01\rangle\right)_{A_1A_2}$$

$$-\sqrt{\frac{|bcd\alpha|^2 + |acd\beta|^2}{|abd\gamma|^2 + |abc\delta|^2}}\left(a^*b^*d^*\gamma e^{-i\varphi_2}|10\rangle + a^*b^*c^*\delta e^{-i\varphi_3}|11\rangle\right)_{A_1A_2}$$

$$|\varphi^4\rangle_{A_1A_2} = k[\sqrt{\frac{|abd\gamma|^2 + |abc\delta|^2}{|bcd\alpha|^2 + |acd\beta|^2}}\left(adc\beta|00\rangle - bcd\alpha e^{-i\varphi_1}|01\rangle\right)_{A_1A_2}$$

$$-\sqrt{\frac{|bcd\alpha|^2 + |acd\beta|^2}{|abd\gamma|^2 + |abc\delta|^2}}\left(-abc\delta e^{-i\varphi_2}|10\rangle + abd\gamma e^{-i\varphi_3}|11\rangle\right)_{A_1A_2} \qquad (40)$$

where $k = \dfrac{1}{\sqrt{|bcd\alpha|^2 + |acd\beta|^2 + |abd\gamma|^2 + |abc\delta|^2}}$.

If Alice's measurement result is $|\varphi^1\rangle_{A_1A_2}$, the state of qubits $B_1, B_2$ has been collapsed into

$$|\varphi^1\rangle_{B_1B_2} = c_2 ab\left(\alpha|00\rangle + \beta e^{i\varphi_1}|01\rangle + \gamma e^{i\varphi_2}|10\rangle + \delta e^{i\varphi_3}|11\rangle\right)_{B_1B_2} \tag{41}$$

Which is exactly the original state $|\varphi\rangle$, and that the successful probability is $k^2|abcd|^2$.

However, when the measurement outcome is $|\varphi^2\rangle_{A_1A_2}, |\varphi^3\rangle_{A_1A_2}, |\varphi^4\rangle_{A_1A_2}$, the remote state preparation can not be successful. Therefore, we can get the probability of successful RSP is

$$k^2|abcd|^2 = \frac{|abcd|^2}{|bcd\alpha|^2 + |acd\beta|^2 + |abd\gamma|^2 + |abc\delta|^2} \tag{42}$$

obviously, if $|a|=|b|=|c|=|d|$, the probability of successful RSP

$$\frac{|abcd|^2}{|bcd\alpha|^2 + |acd\beta|^2 + |abd\gamma|^2 + |abc\delta|^2} = \frac{1}{4}, \tag{43}$$

i.e. the probability of successful RSP is 1/4 with maximally entangled state quantum channel.

**4. Conclusions**

In summary, we have presented a protocol for remote preparation of a single-qubit state and two-particle state with real coefficients and complex coefficients, respectively, which using non-maximally entangled states as the quantum channel. In the two cases the parameters of measurement basis depend on not only the state of preparation but also the state of quantum channel. Because of constructing a set of appropriate mutually orthogonal measurement basis, the success probability is improved. We also calculated the probability of success of the RSP scheme in general and some particular cases. Furthermore, we pointed out that the probability of success may also depend on the sender's choice of measurement basis. If the appropriate measurement is chosen, remote state preparation can be achieved the maximum probability.

**Acknowledgements**
This work was supported by the Natural Science Foundation of Shaanxi Province of China (Grant No. 2013JM1009).